\documentclass[sigconf,nonacm]{acmart}
\usepackage{amsmath, amsthm}
\usepackage{mdframed}
\usepackage{todonotes}

\newtheoremstyle{bigidea}
  {3pt} 
  {3pt} 
  {\itshape} 
  {} 
  {\bfseries} 
  {.} 
  { } 
  {} 

\theoremstyle{bigidea}
\newmdtheoremenv[%
  linecolor=black,
  linewidth=1pt,
  linecolor=black,
  innerleftmargin=10pt,
  innerrightmargin=10pt,
  innertopmargin=6pt,
  innerbottommargin=6pt,
  skipabove=6pt,
  skipbelow=6pt
]{bigidea}[theorem]{Key Aspect}

\AtBeginDocument{%
  }

\setcopyright{acmlicensed}
\copyrightyear{2018}
\acmYear{2018}
\acmDOI{XXXXXXX.XXXXXXX}

\acmConference[Conference acronym 'XX]{Make sure to enter the correct
  conference title from your rights confirmation emai}{June 03--05,
  2018}{Woodstock, NY}
\acmISBN{978-1-4503-XXXX-X/18/06}

\begin{document}

\title{The Role of Generative AI in Software Student CollaborAItion}









\author{Natalie Kiesler}
\affiliation{%
  \institution{Nuremberg Tech}
  \city{Nuremberg}
  \country{Germany}}
\email{natalie.kiesler@th-nuernberg.de}
\orcid{0000-0002-6843-2729}

\author{Jacqueline Smith}
\affiliation{%
  \institution{University of Toronto}
  \city{Toronto}
  \state{Ontario}
  \country{Canada}}
\email{jsmith@cs.toronto.edu}
\orcid{0009-0007-7945-947X}

\author{Juho Leinonen}
\affiliation{%
  \institution{Aalto University}
  \city{Espoo}
  \country{Finland}}
\email{juho.2.leinonen@aalto.fi}
\orcid{0000-0001-6829-9449}

\author{Armando Fox}
\affiliation{%
  \institution{University of California, Berkeley}
  \city{Berkeley}
  \state{California}
  \country{USA}}
\email{fox@berkeley.edu}
\orcid{0000-0002-6096-4931}

\author{Stephen MacNeil}
\affiliation{%
  \institution{Temple University}
  \city{Philadelphia}
  \state{Pennsylvania}
  \country{USA}}
\email{stephen.macneil@temple.edu}
\orcid{0000-0003-2781-6619}

\author{Petri Ihantola}
\affiliation{%
  \institution{University of Jyväskylä}
  \city{Jyväskylä}
  \country{Finland}}
\email{petri.j.ihantola@jyu.fi}
\orcid{0000-0003-1197-7266}

\renewcommand{\shortauthors}{Kiesler et al.}

\begin{abstract}
Collaboration is a crucial part of computing education. The increase in AI capabilities over the last couple of years is bound to profoundly affect all aspects of systems and software engineering, including collaboration. In this position paper, we consider a scenario where AI agents would be able to take on any role in collaborative processes in computing education. We outline these roles, the activities and group dynamics that software development currently include, and discuss if and in what way AI could facilitate these roles and activities. The goal of our work is to envision and critically examine potential futures. We present scenarios suggesting how AI can be integrated into existing collaborations. These are contrasted by design fictions that help demonstrate the new possibilities and challenges for computing education in the AI era. 
\end{abstract}



\begin{teaserfigure}
\centering
  \includegraphics[width=0.85\textwidth]{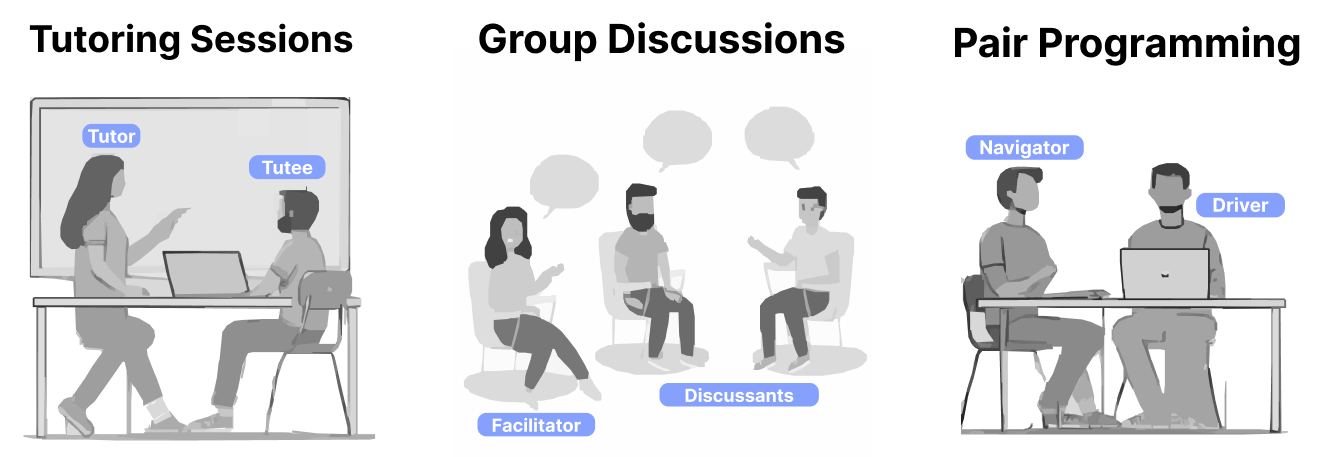}
  \caption{There are a variety of roles that social bots can play in classroom settings. Here social roles are highlighted across three learning contexts including one-on-one tutoring, group discussions, and pair programming.}
  \Description{TODO}
  \label{fig:teaser}
\end{teaserfigure}

\maketitle

\section{Introduction}




By 2030, we will see a generation of college students for whom generative AI (GenAI) has always existed as part of their educational experience. This, and the rapid development of technology, will change teaching and learning. The effects of GenAI on education are likely to be as large as, or even larger than, the effects that the Internet has had since the 1990s (see e.g.~\cite{gouseti2013overview}). 

Khan \cite{khan2024} has proposed an inspiring vision of how AI could help realize personalized individual tutors for every learner. Complementing this, an expert panel from 2020~\cite{roschelle2020ai} draws a scenario where ``AI supports orchestration of the multiple types of activities, learning partners, and interaction patterns that can enrich a classroom''. We believe the possibilities are even broader, and to help think about them, we propose a thought experiment that not only accommodates emerging practices and visions but also suggests new use cases in education that (to the best of our knowledge) have not yet been explored.



Specifically, we focus on the many roles of \emph{collaboration} in Computer Science (CS) education, both because it has been identified as an important component of social constructivist learning theory~\cite{vygotsky1962thought}, and because it is a crucial disposition in the practice of systems and software engineering~\cite{whitehead2007collaboration,mistrik2010collaborative,CC2020,raj2021professional}. 
Taxonomies of collaboration define particular \emph{roles} within a collaborative work scenario. We postulate that in the near future, \emph{AI Agents} will be capable of assuming any given role (e.g., facilitator, mentor, assistant, peer) in any given type of collaboration~\cite{myller2009extending}. 

To keep our focus, we concentrate on systems and software engineering rather than theoretical computer science. In addition, we purposefully do not discuss some issues that current generative AI models have to keep the arguments in the paper focused on the bigger picture (e.g., students cheating and plagiarism~\cite{zastudil2023generative}, copyright issues of training data, biases~\cite{prather2023wgfullreport}). 

The \textbf{goal} is to present key aspects for considering which roles and use cases for AI agents might be desirable, and which ones might be perilous. We are thus creating a framework for thinking about what could happen.
Yet, we are NOT proposing to implement certain use cases. 
Our \textbf{contribution} is to discuss both existing and possible future use cases for AI agents and to identify some potential pitfalls and obstacles in moving forward. Envisioning and critically examining these use cases has implications for educators, tool creators, and beyond.

\section{Why Focus On Collaboration?}

In computing education, collaboration is both an important ingredient in student learning and a desirable competency in itself.  
For example, collaboration is a central component of software development, where teams work together to overcome human limitations and achieve complex results while working at a high level of abstraction~\cite{whitehead2007collaboration,mistrik2010collaborative}. It is therefore not surprising that recent curricula recommendations such as the CC2020~\cite{CC2020} address students' abilities to successfully interact, communicate, and collaborate with others, e.g., in the form of the dispositions \textit{collaborative} and \textit{responsive}. 
Being a team player, and successfully working with others (e.g., team members, leaders, customers, etc.) is also an expectation in industry~\cite{kiesler2024industry}. For this reason, students will likely play different software engineering roles throughout their studies -- developer,  tester,  manager, scrum master,  product owner, and so on. 
Even if AI agents were able to assume most of the roles on a software development team, humans would still need to be able to collaborate with the AI agent(s), for example, to convey what the human wants the AI agent to do. 
Hence we expect collaboration skills to be at least as important in the future as they are today, if not more important.


In collaboration, actors generally work together towards a shared goal, and many taxonomies of collaboration exist~\cite{griffiths2021together,ostergaard2009development,golovchinsky2009taxonomy,nunamaker2001framework,patel2012factors}. In this paper, we take a broad view where collaborators are not always equal in standing such as when teachers and students or students and teaching assistants work together collaboratively. 
Based on social constructivism theory~\cite{vygotsky1962thought}, collaboration tends to improve learning by providing social forms and processes~\cite{howard2001collaborative,kim2001social}. Thus, it would be desirable to maximize collaboration in favor of mere interaction (which does not require a shared goal). 

While in the past, some educational activities have been less collaborative due to technological limitations, we argue that with capable AI agents, some current teaching and learning activities can be transformed from interaction to collaboration. For example, it would be infeasible to have private one-to-one lectures between a human teacher and a human student due to resource constraints, but this is not the case for capable AI agents advising students. Such agents are available 24/7.




\section{Roles and Group Dynamics in Collaboration}

Many of the collaboration taxonomies mentioned above have (at least) two elements in common: 
(a) Each principal or agent in the collaboration takes on a particular \emph{role}, (b) the collaboration can involve one or more \emph{media}, and several \emph{modes}.

For example, a software project course might include several of the following configurations of collaboration:

\begin{itemize}
    \item\textbf{Team meeting.} Students working together on a project meet face-to-face or virtually. Each student takes on the role of a peer. The mode of collaboration is synchronous discussion. In virtual scenarios, respective software and hardware is required as a media.
    \item\textbf{Meeting with project manager/instructor.}  The student team meets with an instructor (``project manager'') who gives them advice on their work. The instructor may take on the role of an evaluator, giving students feedback on the quality of the work; or they may be a facilitator, ensuring that the (less-experienced) students keep the meeting on track.
    \item\textbf{Meeting to resolve or de-escalate a conflict.} An instructor or TA calls a meeting to mediate conflict within the team, 
    whether arising from a difference of technical opinion or a negative power dynamic (see below). The instructor or TA takes the role of a mediator.
    \item\textbf{Pair programming.} Students may pair-program~\cite{hanks2011pair} locally or remotely as peers. The medium is, for example, a collaboratively-authored file of code.
    \item\textbf{Peer instruction.} Students may engage in collaborative quizzes, in-class discussions, or learn by teaching each other~\cite{crouch2001peer,porter2011peer,kothiyal2013effect,grzega2008didactic}. 
\end{itemize}

We will not be surprised to see AI agents interpolated into all of the above roles, and indeed some are already being explored (e.g., using AI in pair programming and peer instruction scenarios).

\begin{bigidea}
Collaboration taxonomies define distinct roles, such as peer or student, instructor, facilitator/mediator, evaluator, and so on. 
It is plausible that AI agents will soon be technically capable of assuming any of these roles. 
\end{bigidea}
\vspace{5pt}

When AI agents become capable of taking on any of the above roles, how will we as educators decide which ones they \emph{should} take on in an educational setting? 
One important consideration is the emergence of various kinds of group dynamics in collaborative work.  
For example, power imbalances may arise between a student and an instructor, or between a student and another student perceived to be more assertive, aggressive, competent, or more fluent in the language of the course. 
Negative dynamics may also arise from implicit or explicit gender bias~\cite{lewis2015equity,shah2019amplifying,stankiewicz2016con} or racial/ethnic bias or stereotypes~\cite{chan2023harms}. 
AI agents could be conditioned to detect, avoid, and/or steer away from such situations.


Human participants can bring positive affect to a discussion, such as inspiration, encouragement in the face of difficulty,
passion for a subject, and so on. They may also share cultural context with other participants that helps to lubricate discussions and avoid or clarify misunderstandings due to communication differences. 
These qualities can be particularly important in educational contexts. We currently do not know how well AI agents can generate such individual perspectives and styles -- and even if so, whose perspectives and styles would they represent?

While AI agents could be conditioned to ``say all the right things'' to mimic human-to-human interactions, we think it is unlikely
that human participants will react in the same way to an AI agent as they would to other humans, assuming they are aware that the agent is an AI.

\begin{bigidea}
The ``emotional stakes'' and affect from interacting with human beings will likely be absent from collaborative interactions with
AI agents. Depending on the context, this absence may manifest as either a positive or negative attribute of the interaction.
\end{bigidea}













\section{Current Uses of AI Agents in Computing}
\label{sec:current}

Even before the GenAI revolution, chatbots have been considered a promising technology to enhance learning. A 2018 systematic literature review (SLR) showed educational effects of chatbots on students regarding their technological skills, social skills, attitude and trust towards technology, self-regulation skills~\cite{winkler2018unleashing}. In 2021, another SLR analyzed how chatbots are applied in education, for which pedagogical roles (e.g., mentoring) they are used, and how they potentially personalize education~\cite{wollny2021are}. The results reveal that chatbots have been attributed a variety of roles, for example, to support learners by teaching content or skills via conversation tasks~\cite{fryer2020supporting}, as a voice assistant during leisure time, or as a pen-pal~\cite{kimna2019astudy}. Another role is that of a mentor, i.e., for scaffolding~\cite{gabrielle2020chatbot}, recommending~\cite{xiao2019who}, informing~\cite{kerly2008calmsystem}, but also to support self-regulation, life, and learning skills~\cite{wollny2021are}. Chatbots were also found to be used as assistants to students, for example, to simplify processes, answer general questions, or make information available~\cite{wollny2021are}.

The development in the field of systems and software engineering has been very similar, with AI agents rapidly gaining popularity since 2019. A SLR by Moguel et al.~\cite{moguel2023bots} from 2023 divides the objectives of AI bots in software engineering (from least common to most common in the literature) into onboarding assistance, testing, design, requirements elicitation, project management, and coding (including code analysis, refactoring, and debugging). According to the same study, challenges typically associated with these tools relate to the performance and capabilities of the tools, high coupling to third-party software, and some developers having a bias against AI agents~\cite{moguel2023bots}.

Notwithstanding the challenges and obstacles in the above work, 
the UI/UX/HCI community has started to systematically explore the more
general question of what new forms of human-computer
interaction are enabled by
GenAI~\cite{bernstein-park-novel-interactions-gen-ai-2023},
and has begun to develop general benchmarks to capture the subjective experience of
human-LLM interaction in a variety of task
types~\cite{lee2024evaluatinghumanlanguagemodelinteraction}. As one example, one recent project populates an entire small village of simulated
residents powered by LLMs, embedding them in a
larger architecture that reifies observation, planning, and
reflection, as a potential way to simulate social
phenomena~\cite{interactive-simulacra-23}.








\section{The Future: CollaborAItion with GenAI}

This position paper aims to envision and explore the potential future of GenAI in computing education, identifying emerging opportunities and challenges to guide future research. Our goal is thus not to prescribe ideal configurations, but to provide a framework that encourages broader thinking about the design space of possibilities and the key considerations that educators need to engage with to maximize benefits while avoiding harm. Accordingly, we introduce some examples of AI agents and how they may be utilized in introductory and advanced courses, as well as in the training of educators or teaching assistants.


\subsection{Introductory Courses}

Many teaching techniques common in introductory courses (e.g., Pair Programming, Think-Pair-Share, etc.) involve putting students in small groups to discuss solving a problem. While these techniques are shown to be effective, there are limitations, for example, when students do not engage with each other. We see a role here for AI agents as conversation facilitators, comparable to an instructor who is walking around the classroom. AI agents could also participate as mock-students, asking questions of their ``peers'' and allowing students to learn through teaching~\cite{cortese2005learning}. 

Pair programming is another widely used teaching approach in both introductory and advanced courses (as well as in industry), with both positive and negative outcomes~\cite{hanks2011pair,lewis2015equity,bowman2019pp}. 
The use of an AI agent as part of a pair could help address negative aspects,
such as the presence of implicit gender bias in pair interactions~\cite{twincode_ese}, while retaining some of the benefits of this method of programming. 

\subsection{Advanced Courses}

In more advanced courses such as software engineering projects, \emph{collaboration} itself is often an important learning goal (even though it is not always explicit).  
Typically, an important component in such courses is team meetings, in which students working together on a project meet face-to-face or virtually. Each student takes on the role of a peer, and the medium of collaboration can be a synchronous discussion.
A recent experience report showed that GenAI tools may help broaden students perspective in the critical thinking process~\cite{grande2024studentperspective}. 

There are several roles usually played by instructors or teaching assistants that AI agents could also assume, relieving the scaling challenges of finding and training staff, e.g., as facilitator, project manager/instructor, mediator, or simply providing personalized feedback and support.

Another interesting role for an AI agent may be that of a substitute team member. When a team member does not fulfill their role for whatever reason, AI agents could play the role of a student teammate, allowing teams to continue to run smoothly even if members drop out or underperform. However, it should be noted that such a substitution has further implications on the replaced student, the student team, and likely the entire class.

\subsection{Teacher and Teaching Assistant Training}

At present, many teachers and teaching assistants (TAs) receive little or no training and feedback on the way they interact with students in one on one or small group settings (e.g., in tutorials or in office hours). An AI agent could join these interactions and provide feedback to the teacher on their interactions with students to help improve their teaching and communication skills. 

We also see opportunities for AI agents in the training of teachers and TAs. A common technique in teacher and TA training is the use of exemplary classroom scenarios, either through case studies or role play. AI agents could be used to play some parts in these scenarios, perhaps allowing more trainees to engage in responding to the scenarios. AI agents may also allow a broader range of scenarios to be explored. Human participants may not be comfortable playing certain parts and AI agents could instead be the actors in more complex scenarios. In addition, AI agents may be used to help educators reflect on the limitations of their own knowledge, perspectives, and positionality, and recognize how their identity is intersected with their educational experience and habits~\cite{freire2020pedagogy,boal2014theatre,schroeter2013way}. 

Beyond teacher and TA training, AI agents could also be used in courses to represent students in course projects or mock research studies that would otherwise have ethical challenges. 
Yet, we note that the involved ethical challenges should be reflected on, and carefully considered in the first place.

\section{Grand Challenges and Future Research}
\label{sec:grandchallenges}


Based on our position that AI agents can and will be integrated into many collaborative learning settings in the future, there are numerous challenges and future research directions to consider. For example, 1) should AI agents be identified as such to students, 2) how might AI agents affect power dynamics~\cite{white2007organization}, 3) how might the identity of students and potential identity or lack of identity of AI agents affect group and classroom dynamics, 4) how much context should be maintained across activities, and 5) how would students who have always interacted with AI agents be different from our current students? In the following sections, we discuss some of the aspects involved in these questions, emphasizing the related challenges and need for future research.



\subsection{Transparency}

If AI agents can assume any role in classroom settings, it is important to better understand students' preferences for these roles and interactions. One area that requires clarity is around transparency. We might assume that students and instructors would be informed and aware that they are interacting with an AI agent. This assumption aligns with The European Union's Artificial Intelligence Act\footnote{\url{https://artificialintelligenceact.eu/high-level-summary/}}
which requires that ``developers and deployers must ensure that end-users are aware that they are interacting with AI (chatbots and deepfakes).'' 
In 2020, Berendt et al.~\cite{berendt2020choice} already highlighted the implications of AI in education for fundamental human rights and freedoms of both teachers and learners. Those comprise, for example, privacy, the protection of personal data, expression and education (presumably a safe space). With capable AI agents, there is an expert pretrained entity in the classroom, which might also collect data. As a consequence, learners may wish to opt out of data collection, or from using AI agents. However, there is currently no global consensus on the issue and few other countries have similar legislation.


\subsection{Power Dynamics and Identity}
\label{sec:powerd}

Prior work has demonstrated how power dynamics play out in computing classrooms, e.g., in pair programming and group discussions~\cite{lewis2015equity,shah2019amplifying,stankiewicz2016con}. 
Rudimentary intelligent systems have also been developed to encourage more equitable group dynamics during collaborative learning activities both in-person~\cite{macneil2019ineqdetect} and online~\cite{stankiewicz2016con}. 

Introducing GenAI agents into classroom settings may further affect power dynamics, as they will likely elevate the position and voice of certain demographics while excluding marginalized groups (with regard to, for example, race, gender, and socio-economic background)~\cite{chan2023harms}. This is due to the inherent limitations of GenAI tools and the underlying models, which have been trained on a limited set of data. The resulting bias and social injustices of GenAI tools~\cite{apiola2024first}, as well as their increasing agency~\cite{chan2024visibility} need to be considered carefully when integrating such tools into collaborative settings. 

There is also evidence that GenAI is changing how much students seek help from their peers~\cite{scholl2024analyzingchatprotocolsnovice,hou2024effects}.
By analyzing the chat protocols of 213 novice learners of programming, Scholl et al.~\cite{scholl2024analyzingchatprotocolsnovice} showed that CS students use GenAI tools in a way that resembles human-to-human interaction, thus using them as a study buddy when they need help. Hou et al.~\cite{hou2024effects} even found that current GenAI tools can lower socio-emotional barriers to seeking help. However, they also provide preliminary evidence that this might also reduce social interactions and discourage students from forming the social support groups on which they previously relied for help.

According to Grande et al.~\cite{grande2018lost}, educators do not just facilitate or present knowledge. They are role models with regard to collaboration, trust, showing care, handling emotions and issues, but also in terms of representation (who they are, which values they represent, etc.). How do AI agents blend into such a complex role and responsibility of an educator? Are they capable of addressing conflicts, or showing care towards a student in a crisis? Can AI agents model self-care, joy, or other human factors and activities for diverse people?  



\subsection{Impacts on Student Community}

Mediation and facilitation can positively impact group dynamics. But more work is needed to understand the impacts of GenAI on student communities. By interacting more frequently with GenAI, students could be missing opportunities to work with, show interest in, and engage with their peers or learn how to interact in social situations. As a consequence, students may not be able to build as many weak and strong ties with their peers, which may lead to a limited support system during their studies.

Some types of AI integration could negatively affect student communities -- for example, if AI is used to replace all peers in group work. These types of effects might be mitigated by having multiple humans play a part in the collaboration when it is feasible. At the same time, it is crucial to ensure that all students have equal opportunities to engage with these human players, and not just AI agents. Hence, we also need to ensure and investigate students' access to all resources -- human and AI, as this does not necessarily apply to all of our students. The same is true for students' prerequisites to successfully utilize AI agents, as mentioned in the previous section on power dynamics and identity (see Section~\ref{sec:powerd}). 

Future research may address the impact of AI agents on students' community building efforts and their ability to build long-lasting social networks with their peers. 


\subsection{Accounting for Context} 

While AI agents are becoming increasingly capable of leveraging context about students, classes, and generated artifacts, challenges remain regarding the extent of their contextual awareness. For instance, students may be aware of deadlines discussed in class or shared via email, but even with classroom microphones and email access, some context will always be missed by these agents. 

Future research could explore ways to expand the context available to AI agents to address contextual gaps that prevent bots from accessing essential information. Additionally, context information may need to be adjusted (i.e., added or even removed) for specific learning activities and situations. 
We have seen such initial efforts related to prompt engineering by tool creators~\cite{denny2023promptly,denny2024prompt} and educational researchers trying to elicit feedback~\cite{azaiz2024feedback,kiesler2023exploring,koutcheme2024evaluating}.
Another common aspect identified in this recent work is the degree of randomness users are confronted with. 

We expect to see more research on this aspect in the next couple of years, e.g., how much do these agents remember, how does the work on a single machine or across multiple spaces affect responses, and how should we define a common ground for learners to succeed in their communication and collaboration?




\subsection{(Lack of) AI Literacy}

As AI agents become more ubiquitous and students become more familiar with the interactions, new challenges may emerge. For example, even if interactions with agents become more natural, we still need to remember to teach the underlying complexities. There is an analogy with the so-called ``digital natives'', who were first assumed to have better digital skills simply because they had never known a world that was not digital -- which later turned out not to be true~\cite{kirschner2017themyths}. For example, many students in CS1 courses today seem to struggle with basic things, e.g., following the file structure or understanding class paths, due to those things not being explicitly taught and modern systems such as smartphones hiding them from the user, and at the same time, their reliance on search and other mechanisms.
Students in the future may be more competent as users applying GenAI tools, but they will not necessarily understand the tools, their architecture, etc. (e.g., how a Google search works, and how GenAI works). It is therefore crucial to inform them about the inherent limitations of AI agents, such as the bias of the training data, its knowledge cutoff date, etc., as we cannot assume AI agents will do so by themselves.

\begin{bigidea}
All of these grand challenges are related to AI agents' limited consciousness and lack of affect,
as well as the limited transparency within the systems by design.
\end{bigidea}

\section{Limitations}

This position paper has some limitations that need to be considered. For example, we assume AI agents to become extremely proficient in assisting learners with individual requests. However, this scenario might not happen. Moreover, we acknowledge and understand that we might see human resistance toward AI agents in learning and teaching contexts. 

Likewise, we refrained from taking a side in the discussion on the future direction of AI agents in collaboration. From our perspective, it is still too early for wild speculations about the distant future.

Finally, it should be noted that the scenarios we discussed only focus on software engineering as part of computing education. It is not the goal of this position paper to make assumptions about education as a whole.


\section{Conclusions}

The main contribution of this position paper is to provide an outline of existing collaborative activities and roles in systems and software education, and to discuss how highly-capable AI agents might affect collaboration in this field. We believe that AI agents will soon be capable of assuming different roles in collaborative activities. It is crucial to carefully consider what effects AI might have on different aspects of collaboration, and how decisions made when integrating AI affect its impact.

GenAI is a potentially transformative technology for our educational system. We must be careful not to fall into the trap of simply substituting it into existing educational scenarios, but rather recognizing that it will enable entirely new modes of learning that were previously impossible or even unthinkable. Identifying the CS concepts that remain foundational, the intellectual vocabulary needed to identify and exploit those new learning modes, and the challenges and opportunities for introduced by doing so, should be on our agenda.

\begin{acks}
This work was initialized at and has benefitted substantially from the Dagstuhl Seminar 24272 ``A Game of Shadows: Effective Mastery Learning in the Age of Ubiquitous AI'' at Schloss Dagstuhl -- Leibniz Center for Informatics.

This research was supported by the Research Council of Finland (Academy Research Fellow grant number 356114), by the California Education Learning Lab, an initiative of the California Governor’s Office of Planning and Research (grant \#00005919), and by the European Union (OSCAR project, Erasmus+ grant number 101132432). Views and opinions expressed are, however, those of the authors only.

We are grateful that ChatGPT was able to help us facilitate the choice of author order.
\end{acks}

\balance
\bibliographystyle{ACM-Reference-Format}
\bibliography{references}


\end{document}